\documentclass[preprintnumbers,amsmath,amssymbm,prd]{revtex4}
\usepackage{epsfig}
\usepackage{graphicx}
\usepackage{amssymb}

\begin{document}
\title{Stationary bound-state scalar configurations supported by rapidly-spinning exotic compact objects}
\author{Shahar Hod}
\affiliation{The Ruppin Academic Center, Emeq Hefer 40250, Israel}
\affiliation{ } \affiliation{The Hadassah Institute, Jerusalem
91010, Israel}
\date{\today}

\begin{abstract}
\ \ \ Some quantum-gravity theories suggest that the absorbing
horizon of a classical black hole should be replaced by a reflective
surface which is located a microscopic distance above the would-be
classical horizon. Instead of an absorbing black hole, the resulting
horizonless spacetime describes a reflective exotic compact object.
Motivated by this intriguing prediction, in the present paper we
explore the physical properties of exotic compact objects which are
linearly coupled to stationary bound-state massive scalar field
configurations. In particular, solving the Klein-Gordon wave
equation for a stationary scalar field of proper mass $\mu$ and
spheroidal harmonic indices $(l,m)$ in the background of a
rapidly-rotating exotic compact object of mass $M$ and angular
momentum $J=Ma$, we derive a compact analytical formula for the {\it
discrete} radii $\{r_{\text{c}}(\mu,l,m,M,a;n)\}$ of the exotic
compact objects which can support the stationary bound-state massive
scalar field configurations. We confirm our analytical results by
direct numerical computations.
\end{abstract}
\bigskip
\maketitle

%]

\section{Introduction}

Black holes in classical theories of gravity describe compact
spacetime regions which are bounded by event horizons with absorbing
boundary conditions. Interestingly, however, some candidate
quantum-gravity models
\cite{eco1,eco2,eco3,eco4,eco5,eco6,eco7,eco8,eco9,eco10,eco11,eco12,Pan}
have recently suggested that quantum effects may prevent the
formation of stable black-hole horizons. These models have put
forward the intriguing idea that, within the framework of a quantum
theory of gravity, horizonless exotic compact objects may serve as
alternatives to the familiar classical black-hole spacetimes
\cite{eco1,eco2,eco3,eco4,eco5,eco6,eco7,eco8,eco9,eco10,eco11,eco12,Pan}.

In a very interesting work, Maggio, Pani, and Ferrari \cite{Pan}
have recently studied numerically the physical properties of
spinning exotic compact objects which are characterized by spacetime
geometries that modify the familiar Kerr metric only at some
microscopic scale around the would-be classical horizon. In
particular, the physical model analyzed in \cite{Pan} assumes that
the absorbing horizon of the classical Kerr black-hole spacetime is
replaced by a slightly larger quantum membrane with reflective
boundary conditions.

The interplay between compact astrophysical objects and fundamental
matter fields has attracted much attention over the years from both
physicists and mathematicians. In particular, recent analytical
\cite{Hodrc} and numerical \cite{Herkr} studies of the
Einstein-Klein-Gordon field equations have revealed the fact that,
within the framework of classical general relativity, spinning Kerr
black holes can support stationary spatially regular bound-state
matter configurations which are made of massive scalar fields. This
fact naturally raises the following physically interesting question:
Can the exotic compact objects of the suggested quantum-gravity
models
\cite{eco1,eco2,eco3,eco4,eco5,eco6,eco7,eco8,eco9,eco10,eco11,eco12,Pan},
like the more familiar classical black holes \cite{Hodrc,Herkr},
support stationary bound-state massive scalar field configurations
in their exterior regions?

In the present paper we shall address this physically intriguing
question by solving the Klein-Gordon wave equation for massive
scalar fields in the background of a rapidly-rotating exotic compact
object. In particular, motivated by the suggested quantum-gravity
model recently studied in \cite{Pan} (see also
\cite{eco1,eco2,eco3,eco4,eco5,eco6,eco7,eco8,eco9,eco10,eco11,eco12}),
we shall use {\it analytical} techniques in order to study the
physical properties of exotic compact objects with reflective
boundary conditions which are linearly coupled to stationary
bound-state massive scalar field configurations. Interestingly, as
we shall explicitly show below, one can derive a remarkably compact
analytical formula for the {\it discrete} radii
$\{r_{\text{c}}(\mu;n)\}$ \cite{Notennn} of the horizonless
rapidly-spinning exotic compact objects which, for a given proper
mass $\mu$ of the external field, can support the stationary
bound-state massive scalar field configurations.

It is important to note that the horizonless spinning configurations
that we shall study in the present paper are characterized by the
dimensionless relation $M\mu=O(m)$, where $M$ is the mass of the
central exotic compact object and $m$ is the azimuthal harmonic
index of the massive scalar field mode. This characteristic relation
implies, in particular, that for astrophysically realistic
black-hole candidates, the relevant massive scalar fields are
ultralight: $\mu\sim10^{-10}-10^{-19}$eV. In this context, it is
worth noting that the physical motivation to consider such
ultralight massive scalar fields is manifold and ranges from
possible dark matter candidate fields to new fundamental bosonic
fields which might appear in suggested extensions of the Standard
Model and in fundamental string theories. In particular, such
ultralight fields naturally appear in the suggested string axiverse
scenario, in theories of exotic dark photons, and in hidden $U(1)$
sectors \cite{Arv1,Arv2,Feng,Hui,Wik}.

Before proceeding, it is worth emphasizing that the composed
spinning-exotic-compact-object-massive-scalar-field configurations
that we shall analyze in the present paper, like the composed
spinning-black-hole-massive-scalar-field configurations studied in
\cite{Hodrc,Herkr}, owe their existence to the intriguing physical
phenomenon of superradiant scattering in rotating spacetimes
\cite{Frid,Bri}. In particular, the stationary field configurations,
whose physical properties will be analyzed below, are characterized
by the critical (marginal) frequency $\omega_{\text{c}}$ for the
superradiant scattering phenomenon of bosonic fields in spinning
spacetimes.

\section{Description of the system}

We shall study analytically the physical properties of massive
scalar field configurations which are linearly coupled to a
rapidly-spinning exotic compact object. Motivated by the suggested
quantum-gravity models
\cite{eco1,eco2,eco3,eco4,eco5,eco6,eco7,eco8,eco9,eco10,eco11,eco12,Pan},
we shall assume that the spacetime of the exotic compact object is
described by a curved geometry that modifies the familiar Kerr
metric only at some microscopic scale around the would-be classical
horizon. In particular, following the interesting work of Maggio,
Pani, and Ferrari \cite{Pan}, we shall consider a reflective
spinning compact object of radius $r_{\text{c}}$, mass $M$, and
angular momentum $J=Ma$ whose exterior spacetime metric is described
by the curved Kerr line element \cite{Chan,Noteman} (we shall use
natural units in which $G=c=\hbar=1$)
\begin{eqnarray}\label{Eq1}
ds^2=-{{\Delta}\over{\rho^2}}(dt-a\sin^2\theta
d\phi)^2+{{\rho^2}\over{\Delta}}dr^2+\rho^2
d\theta^2+{{\sin^2\theta}\over{\rho^2}}\big[a
dt-(r^2+a^2)d\phi\big]^2\ \ \ \text{for}\ \ \ \ r>r_{\text{c}}\
\end{eqnarray}
with $\Delta\equiv r^2-2Mr+a^2$ and $\rho^2\equiv
r^2+a^2\cos^2\theta$. Here $(t,r,\theta,\phi)$ are the familiar
Boyer-Lindquist spacetime coordinates. The radius of the would-be
classical horizon is given by
\begin{equation}\label{Eq2}
r_+=M+(M^2-a^2)^{1/2}\  .
\end{equation}
Following the intriguing quantum-gravity models
\cite{eco1,eco2,eco3,eco4,eco5,eco6,eco7,eco8,eco9,eco10,eco11,eco12,Pan},
which predict the occurrence of quantum corrections to the curved
spacetime only at some microscopic scale around the would-be
classical horizon, we shall assume that the radius $r_{\text{c}}$ of
the reflective quantum membrane is characterized by the strong
inequality
\begin{equation}\label{Eq3}
x_{\text{c}}\equiv {{r_{\text{c}}-r_+}\over{r_+}}\ll1\  .
\end{equation}
It is important to emphasize that the assumption made in \cite{Pan},
that the exterior spacetime region of the spinning exotic compact
object is described by the Kerr metric (\ref{Eq1}), is a non-trivial
one. As emphasized in \cite{Pan} (see also Refs.
\cite{kr1,kr2,kr3}), this assumption is expected to be valid in the
physically interesting regime $x_{\text{c}}\ll1$ of small quantum
corrections.

The Klein-Gordon wave equation \cite{Teuk,Stro}
\begin{equation}\label{Eq4}
(\nabla^\nu\nabla_{\nu}-\mu^2)\Psi=0\
\end{equation}
governs the dynamics of the scalar field in the curved spacetime of
the exotic compact object, where $\mu$ is the proper mass of the
linearized field \cite{Noteunm}. Substituting into (\ref{Eq4}) the
mathematical decomposition \cite{Teuk,Stro,Notedec}
\begin{equation}\label{Eq5}
\Psi(t,r,\theta,\phi)=\sum_{l,m}e^{im\phi}{S_{lm}}(\theta;a\sqrt{\mu^2-\omega^2}){R_{lm}}(r;M,a,\mu,\omega)e^{-i\omega
t}\
\end{equation}
for the eigenfunction $\Psi$ of the massive scalar field, and using
the metric components (\ref{Eq1}) which characterize the exterior
curved spacetime of the exotic compact object, one finds that the
radial eigenfunction of the massive scalar field satisfies the
ordinary differential equation \cite{Teuk,Stro}
\begin{equation}\label{Eq6}
\Delta{{d}
\over{dr}}\Big(\Delta{{dR_{lm}}\over{dr}}\Big)+\Big\{[\omega(r^2+a^2)-ma]^2
+\Delta[2ma\omega-\mu^2(r^2+a^2)-K_{lm}]\Big\}R_{lm}=0\ ,
\end{equation}
where the angular eigenvalues $\{K_{lm}(a\sqrt{\mu^2-\omega^2})\}$
are determined by the characteristic angular differential equation
\cite{Heun,Fiz1,Teuk,Abram,Stro,Hodasy,Hodpp}
\begin{eqnarray}\label{Eq7}
{1\over {\sin\theta}}{{d}\over{\theta}}\Big(\sin\theta {{d
S_{lm}}\over{d\theta}}\Big) +\Big[K_{lm}+a^2(\mu^2-\omega^2)
-a^2(\mu^2-\omega^2)\cos^2\theta-{{m^2}\over{\sin^2\theta}}\Big]S_{lm}=0\
\end{eqnarray}
with the physically motivated boundary conditions of regularity at
the two angular poles $\theta=0$ and $\theta=\pi$. It is worth
noting that the characteristic eigenvalues of the spheroidal angular
equation (\ref{Eq7}) can be expanded in the form
$K_{lm}+a^2(\mu^2-\omega^2)=l(l+1)+\sum_{k=1}^{\infty}c_k
a^{2k}(\mu^2-\omega^2)^k$, where the expansion coefficients
$\{c_k(l,m)\}$ are given in \cite{Abram}.

The stationary bound-state configurations of the spatially regular
massive scalar fields in the curved spacetime of the spinning exotic
compact object are characterized by exponentially decaying
(normalizable) radial eigenfunctions at spatial infinity
\cite{Notemas} [for brevity, we shall henceforth omit the angular
harmonic indices $(l,m)$ which characterize the spatially regular
massive scalar field mode]:
\begin{equation}\label{Eq8}
R(r\to\infty)\sim {{1}\over{r}}e^{-\sqrt{\mu^2-\omega^2}r}\ \ \ \
\text{with}\ \ \ \ \omega^2<\mu^2\  .
\end{equation}
In addition, following the quantum-gravity model studied in
\cite{Pan}, we shall assume that the exotic compact object is
characterized by a reflecting surface which is located a microscopic
distance [see Eq. (\ref{Eq3})] above the would-be classical horizon.
In particular, we shall consider two types of inner boundary
conditions \cite{Pan}:
\begin{equation}\label{Eq9}
\begin{cases}
R(r=r_{\text{c}})=0 &\ \ \ \ \text{Dirichlet B. C.}\ ; \\
dR(r=r_{\text{c}})/dr=0 &\ \ \ \ \text{Neumann B. C.}\ \ .
\end{cases}
\end{equation}

The set of equations (\ref{Eq6})-(\ref{Eq9}) determines the {\it
discrete} spectrum of radii $\{r_{\text{c}}(\mu,l,m,M,a;n)\}$ of the
spinning exotic compact objects which can support the spatially
regular stationary bound-state massive scalar field configurations.
Interestingly, in the next section we shall explicitly prove that,
for rapidly-rotating horizonless compact objects, this
characteristic discrete spectrum of supporting radii can be
determined {\it analytically}.

\section{The resonance conditions of the composed spinning-exotic-compact-object-massive-scalar-field
configurations}

In the present section we shall analyze the set of equations
(\ref{Eq6})-(\ref{Eq9}) which determine the bound-state resonances
of the massive scalar fields in the curved background of the
spinning horizonless exotic compact object. In particular, below we
shall use {\it analytical} techniques in order to derive remarkably
compact resonance conditions for the {\it discrete} radii
$\{r^{\text{Dirichlet}}_{\text{c}}(\mu,l,m,M,a;n)\}$ and
$\{r^{\text{Neumann}}_{\text{c}}(\mu,l,m,M,a;n)\}$ of the
rapidly-spinning exotic compact objects which can support the
spatially regular stationary bound-state massive scalar field
configurations.

It is convenient to use the dimensionless physical parameters
\cite{Teuk,Stro}
\begin{equation}\label{Eq10}
x\equiv {{r-r_+}\over {r_+}}\ \ \ ;\ \ \
\tau\equiv{{r_+-r_-}\over{r_+}}\ \ \ ;\ \ \ k\equiv 2\omega r_+\ \ \
;\ \ \ \bar\mu\equiv M\mu\ \ \ ;\ \ \ \omega_{\text{c}}r_+\equiv
{{ma}\over{2M}}\ \ \ ;\ \ \
\varpi\equiv{{2M(\omega-\omega_{\text{c}})}\over{\tau}}\ ,
\end{equation}
in terms of which the scalar radial equation (\ref{Eq6}) takes the
form \cite{Hodcm,Stro}
\begin{equation}\label{Eq11}
x(x+\tau){{d^2R}\over{dx^2}}+(2x+\tau){{dR}\over{dx}}+UR=0\  ,
\end{equation}
where
\begin{equation}\label{Eq12}
U(x)={{(\omega
r_+x^2+kx+\varpi\tau)^2}\over{x(x+\tau)}}-K+2ma\omega-\mu^2[r^2_+(1+x)^2+a^2]\
.
\end{equation}

Interestingly, and most importantly for our analysis, the radial
differential equation (\ref{Eq11}) is amenable to an analytical
treatment in the spatial region
\begin{equation}\label{Eq13}
x\geq x_{\text{c}}\gg \tau\times\text{max}(1,\varpi)\  .
\end{equation}
Note that the strong inequalities (\ref{Eq3}) and (\ref{Eq13}) can
be satisfied simultaneously in the regime
\begin{equation}\label{Eq14}
\tau\times\text{max}(1,\varpi)\ll1\  ,
\end{equation}
in which case one may approximate the massive scalar equation
(\ref{Eq11}) by \cite{Hodcm,Stro}
\begin{equation}\label{Eq15}
x^2{{d^2R}\over{dx^2}}+2x{{dR}\over{dx}}+\bar U R=0\  ,
\end{equation}
where
\begin{equation}\label{Eq16}
\bar
U(x)=[(m^2/4-{\bar\mu}^2)x^2+(m^2-2{\bar\mu}^2)x+(-K+2m^2-2{\bar\mu}^2)]\cdot[1+O(\tau,\tau\varpi)]\
.
%$\bar U=(\omega r_+x+k)^2-K+2ma\omega-\mu^2[r^2_+(1+x)^2+a^2]$.
\end{equation}
It should be emphasized that the $\tau\ll1$ regime (\ref{Eq14})
corresponds to rapidly-spinning exotic compact objects. In addition,
it is worth noting that the physical significance of the field
frequency $\omega=\omega_{\text{c}}$ [which corresponds to
$\varpi=0$, see Eq. (\ref{Eq10})] stems from the fact that, for
classical spinning black-hole spacetimes, this unique resonant
frequency characterizes the composed stationary
black-hole-massive-scalar-field configurations studied in
\cite{Hodrc,Herkr}. Note, in particular, that in (\ref{Eq16}) we
have used the dimensionless relation $M\omega={1\over
2}m\cdot[1+O(\tau,\tau\varpi)]$ in the regime (\ref{Eq14}) of
rapidly-spinning exotic compact objects and field frequencies that
lie in the vicinity of the critical superradiant frequency
$\omega=\omega_{\text{c}}$ (or equivalently, in the vicinity of
$\varpi=0$).

The general mathematical solution of the radial differential
equation (\ref{Eq15}) can be expressed in terms of the
%confluent hypergeometric functions.
familiar Whittaker functions \cite{Abram,Morse}. In particular,
defining the dimensionless variables
\begin{equation}\label{Eq17}
\epsilon\equiv \sqrt{{\bar\mu}^2-m^2/4}\ \ \ \ ; \ \ \ \
\kappa\equiv {{m^2/2-{\bar\mu}^2}\over{\epsilon}}\ \ \ \ ; \ \ \ \
\delta^2\equiv -K-{1\over 4}+2(m^2-{\bar\mu}^2)\  ,
\end{equation}
one finds \cite{Abram,Morse,Hodcm,Notehc}
\begin{equation}\label{Eq18}
R(x)=x^{-1}\big[A\cdot W_{\kappa,i\delta}(2\epsilon x)+B\cdot
M_{\kappa,i\delta}(2\epsilon x)\big]\  ,
%R(x)=x^{-{1\over 2}+i\delta}e^{-\epsilon x}\big[A\cdot U({1\over
%2}+i\delta-\kappa,1+2i\delta,2\epsilon x)+B\cdot M({1\over
%2}+i\delta-\kappa,1+2i\delta,2\epsilon x)\big]\  ,
\end{equation}
where $\{A,B\}$ are normalization constants. We shall henceforth
assume that \cite{Notedp}
\begin{equation}\label{Eq19}
\{\delta,\kappa\}\in \mathbb{R}\  .
\end{equation}
Note that the assumption $\kappa\in \mathbb{R}$ corresponds to
bound-state resonances of the massive scalar fields with
$\omega^2<\mu^2$ [see Eqs. (\ref{Eq8}) and (\ref{Eq17})].

The asymptotic spatial behavior of the radial scalar function
(\ref{Eq18}) is given by \cite{Noteasa}
\begin{equation}\label{Eq20}
R(x\to\infty)=A\cdot x^{-1+\kappa}e^{-\epsilon x}+B\cdot
{{\Gamma(1+2i\delta)}\over{\Gamma({1\over
2}+i\delta-\kappa)}}x^{-1-\kappa}e^{\epsilon x}\  .
\end{equation}
Taking cognizance of the asymptotic boundary condition (\ref{Eq8}),
which characterizes the spatially regular bound-state (normalizable)
configurations of the massive scalar fields, one concludes that the
coefficient of the exploding exponent in the asymptotic radial
expression (\ref{Eq20}) should vanish:
\begin{equation}\label{Eq21}
B=0\  .
\end{equation}
We therefore find that the stationary bound-state resonances of the
massive scalar fields in the horizonless curved spacetime of the
rapidly-spinning exotic compact object are characterized by the
spatially regular radial eigenfunction
\begin{equation}\label{Eq22}
R(x)=A\cdot x^{-1} W_{\kappa,i\delta}(2\epsilon x)\  ,
%R(x)=A\cdot x^{-{1\over 2}+i\delta}e^{-\epsilon x} U({1\over
%2}+i\delta-\kappa,1+2i\delta,2\epsilon x)\  ,
\end{equation}
where $W_{\kappa,\beta}(z)$ is the familiar Whittaker function
%$U(a,b,z)$ is the familiar confluent hypergeometric function
of the second kind \cite{Abram}.

Taking cognizance of the inner boundary conditions (\ref{Eq9}) at
the reflective surface $x=x_{\text{c}}$ of the exotic compact
object, together with the characteristic radial eigenfunction
(\ref{Eq22}) of the stationary bound-state massive scalar field
configurations, one obtains the remarkably compact resonance
conditions
\begin{equation}\label{Eq23}
%U({1\over 2}+i\delta-\kappa,1+2i\delta,2\epsilon x_{\text{c}})=0
W_{\kappa,i\delta}(2\epsilon x_{\text{c}})=0\ \ \ \ \text{for}\ \ \
\ \text{Dirichlet B. C.}\
\end{equation}
and
\begin{equation}\label{Eq24}
{{d}\over{dx}}[x^{-1}W_{\kappa,i\delta}(2\epsilon
x)]_{x=x_{\text{c}}}=0\ \ \ \ \text{for}\ \ \ \ \text{Neumann B.
C.}\
\end{equation}
for the composed
exotic-compact-object-linearized-massive-scalar-field
configurations. The analytically derived resonance conditions
(\ref{Eq23}) and (\ref{Eq24}) determine the dimensionless {\it
discrete} radii $\{x_{\text{c}}(\mu,l,m,M,a;n)\}$ of the
rapidly-spinning exotic compact objects which can support the
stationary bound-state spatially regular massive scalar field
configurations.

In the next section we shall use analytical techniques in order to
prove that the resonance conditions (\ref{Eq23}) and (\ref{Eq24})
for the characteristic dimensionless radii of the central exotic
compact objects can only be satisfied in the bounded radial regime
\begin{equation}\label{Eq25}
\epsilon x_{\text{c}}<\kappa+\sqrt{\kappa^2+\delta^2}\ .
\end{equation}

\section{Upper bound on the radii of the supporting exotic compact objects}

In the present section we shall use a simple analytical argument in
order to derive a remarkably compact upper bound on the
characteristic dimensionless radii $\{x_{\text{c}}(\mu,l,m,M,a;n)\}$
of the rapidly-spinning exotic compact objects which can support the
stationary bound-state massive scalar field configurations.

It proves useful to define the new radial function
\begin{equation}\label{Eq26}
R=x^{\gamma}\Phi\ \ \ \text{with}\ \ \ \gamma\geq -1\  ,
\end{equation}
in terms of which the radial differential equation (\ref{Eq15}) can
be written in the form
\begin{equation}\label{Eq27}
x^2{{d^2\Phi}\over{dx^2}}+2(1+\gamma)x{{d\Phi}\over{dx}}+[{\bar
U}(x)+\gamma(\gamma+1)]\Phi=0\ .
\end{equation}
Taking cognizance of the boundary conditions (\ref{Eq8}) and
(\ref{Eq9}), one concludes that the eigenfunction $\Phi(x)$, which
characterizes the spatially regular stationary bound-state
configurations of the massive scalar fields in the background of the
exotic compact object, must have (at least) one inflection point,
$x=x_{\text{in}}$, in the radial interval
\begin{equation}\label{Eq28}
x_{\text{in}}\in (x_{\text{c}},\infty)\
\end{equation}
which is characterized by the functional relations
\begin{equation}\label{Eq29}
\{\Phi\cdot{{d\Phi}\over{dx}}<0\ \ \ \text{and}\ \ \
{{d^2\Phi}\over{dx^2}}=0\}\ \ \ \ \text{for}\ \ \ \ x=x_{\text{in}}\
.
\end{equation}

Substituting the characteristic relations (\ref{Eq29}) into the
radial scalar equation (\ref{Eq27}), one concludes that the composed
exotic-compact-object-linearized-massive-scalar-field configurations
are characterized by the inequality
\begin{equation}\label{Eq30}
{\bar U}(x_{\text{in}})+\gamma(\gamma+1)>0\  ,
\end{equation}
which implies [see Eq. (\ref{Eq16})]
\begin{equation}\label{Eq31}
\epsilon^2\cdot x^2_{\text{in}}-2\epsilon\kappa\cdot
x_{\text{in}}-[\delta^2+{1\over4}+\gamma(\gamma+1)]<0\ .
\end{equation}
Taking cognizance of (\ref{Eq28}) and (\ref{Eq31}), one finds the
upper bound
\begin{equation}\label{Eq32}
x_{\text{c}}<{{\kappa+\sqrt{\kappa^2+\delta^2+{1\over4}+\gamma(\gamma+1)}}\over{\epsilon}}\
.
\end{equation}
The strongest upper bound on the dimensionless radius $x_{\text{c}}$
can be obtained by minimizing the r.h.s of (\ref{Eq32}). In
particular, the term $\gamma(\gamma+1)$ is minimized for
$\gamma=-1/2$, in which case one finds from (\ref{Eq32}) the
characteristic upper bound
\begin{equation}\label{Eq33}
x_{\text{c}}<{{\kappa+\sqrt{\kappa^2+\delta^2}}\over{\epsilon}}\ .
\end{equation}
on the dimensionless radii of the rapidly-spinning exotic compact
objects which can support the stationary bound-state massive scalar
field configurations.

\section{The characteristic resonance spectra of the stationary composed
exotic-compact-object-linearized-massive-scalar-field
configurations}

Interestingly, as we shall now prove explicitly, the resonance
equations (\ref{Eq23}) and (\ref{Eq24}), which determine the
discrete radii $\{x_{\text{c}}(\mu,l,m,M,a;n)\}$ of the supporting
exotic compact objects, can be solved {\it analytically} in the
physically interesting regime
\begin{equation}\label{Eq34}
x_{\text{c}}\ll1\  .
\end{equation}
It is worth emphasizing again that the strong inequality
(\ref{Eq34}) characterizes exotic compact objects whose quantum
reflective surfaces are located in the radial vicinity of the
would-be classical black-hole horizons [see Eqs. (\ref{Eq2}) and
(\ref{Eq3})]. In particular, the small-$x_{\text{c}}$ regime
(\ref{Eq34}) corresponds to the physically interesting model of
spinning exotic compact objects recently studied numerically in
\cite{Pan} (see also
\cite{eco1,eco2,eco3,eco4,eco5,eco6,eco7,eco8,eco9,eco10,eco11,eco12}).

Using Eqs. 13.1.3, 13.1.33, and 13.5.5 of \cite{Abram}, one can
express the (Dirichlet) resonance condition (\ref{Eq23}) in the form
\begin{equation}\label{Eq35}
(2\epsilon
x_{\text{c}})^{2i\delta}={{\Gamma(1+2i\delta)\Gamma({1\over
2}-i\delta-\kappa)}\over{\Gamma(1-2i\delta)\Gamma({1\over
2}+i\delta-\kappa)}}\ \ \ \ \ \text{for}\ \ \ \ x_{\text{c}}\ll1\  .
\end{equation}
Likewise, using Eqs. 13.1.3, 13.1.33, 13.4.33, and 13.5.5 of
\cite{Abram}, one can express the (Neumann) resonance condition
(\ref{Eq24}) in the form
\begin{equation}\label{Eq36}
(2\epsilon
x_{\text{c}})^{2i\delta}={{\Gamma(2+2i\delta)\Gamma({1\over
2}-i\delta-\kappa)}\over{\Gamma(2-2i\delta)\Gamma({1\over
2}+i\delta-\kappa)}}\ \ \ \ \ \text{for}\ \ \ \ x_{\text{c}}\ll1\  .
\end{equation}
From Eqs. (\ref{Eq35}) and (\ref{Eq36}) one obtains respectively the
analytical formulas \cite{Notenn}
\begin{equation}\label{Eq37}
x^{\text{Dirichlet}}_{\text{c}}(n)={{e^{-\pi
n/\delta}}\over{2\epsilon}}\Big[{{\Gamma(1+2i\delta)\Gamma({1\over
2}-i\delta-\kappa)}\over{\Gamma(1-2i\delta)\Gamma({1\over
2}+i\delta-\kappa)}}\Big]^{1/2i\delta}\ \ \ ; \ \ \ n\in\mathbb{Z}
\end{equation}
and
\begin{equation}\label{Eq38}
x^{\text{Neumann}}_{\text{c}}(n)={{e^{-\pi
n/\delta}}\over{2\epsilon}}\Big[{{\Gamma(2+2i\delta)\Gamma({1\over
2}-i\delta-\kappa)}\over{\Gamma(2-2i\delta)\Gamma({1\over
2}+i\delta-\kappa)}}\Big]^{1/2i\delta}\ \ \ ; \ \ \ n\in\mathbb{Z}\
\end{equation}
for the dimensionless {\it discrete} radii of the horizonless
rapidly-spinning exotic compact objects which can support the
stationary bound-state massive scalar field configurations
\cite{Notekp0}.

It is worth noting that, taking cognizance of Eq. 6.1.23 of
\cite{Abram}, one deduces that
$\Gamma(1+2i\delta)/\Gamma(1-2i\delta)=e^{i\phi_1}$,
$\Gamma(2+2i\delta)/\Gamma(2-2i\delta)=e^{i\phi_2}$, and
$\Gamma(1/2-i\delta-\kappa)/\Gamma(1/2+i\delta-\kappa)=e^{i\phi_3}$
for $\{\delta,\kappa\}\in \mathbb{R}$ [see Eq. (\ref{Eq19})], where
$\{\phi_1,\phi_2,\phi_3\}\in\mathbb{R}$. These relations imply that
$\{x^{\text{Dirichlet}}_{\text{c}}(n),x^{\text{Neumann}}_{\text{c}}(n)\}\in\mathbb{R}$.

\section{The eikonal large-mass $M\mu\gg1$ regime}

In the present section we shall show that the analytically derived
resonance spectra (\ref{Eq37}) and (\ref{Eq38}), which characterize
the composed exotic-compact-object-linearized-massive-scalar-field
configurations, can be further simplified in the asymptotic eikonal
regime
\begin{equation}\label{Eq39}
m\gg1\
\end{equation}
of large field masses \cite{Notelmlm}.

In the regime (\ref{Eq39}) one finds the compact asymptotic
expression \cite{Hodpp,Notewlm}
\begin{equation}\label{Eq40}
K_{mm}={5\over
4}m^2-{\bar\mu}^2+\sqrt{{3\over4}m^2+{\bar\mu}^2}+O(1)\ ,
\end{equation}
which implies [see Eq. (\ref{Eq17})]
\begin{equation}\label{Eq41}
\delta=\sqrt{{3\over
4}m^2-{\bar\mu}^2-\sqrt{{3\over4}m^2+{\bar\mu}^2}}\cdot[1+O(\tau,\tau\varpi)]\
.
\end{equation}
In addition, in the large-$m$ regime one may use the asymptotic
approximations \cite{Abram}
\begin{equation}\label{Eq42}
{{\Gamma(2i\delta)}\over{\Gamma(-2i\delta)}}=i\Big({{2\delta}\over{e}}\Big)^{4i\delta}\cdot[1+O(m^{-1})]\
\end{equation}
and
\begin{equation}\label{Eq43}
{{\Gamma({1\over 2}-i\delta-\kappa)}\over{\Gamma({1\over
2}+i\delta-\kappa)}}=e^{2i\delta}(\delta^2+\kappa^2)^{-i\delta}e^{2i\pi\kappa(1-\theta)}[1+O(m^{-1})]\
\ \ \text{where}\ \ \ \theta\equiv \pi^{-1}\arctan(\delta/\kappa)\ .
\end{equation}

Substituting (\ref{Eq42}) and (\ref{Eq43}) into the resonant spectra
(\ref{Eq37}) and (\ref{Eq38}), one finds respectively the simplified
expressions \cite{Notegmf}
\begin{equation}\label{Eq44}
x^{\text{Dirichlet}}_{\text{c}}(n)={{2e^{\pi[\kappa(1-\theta)-1/4-n]/\delta}}
\over{e\sqrt{\delta^2+\kappa^2}\epsilon}}\
\ \ ; \ \ \ n\in\mathbb{Z}
\end{equation}
and
\begin{equation}\label{Eq45}
x^{\text{Neumann}}_{\text{c}}(n)={{2e^{\pi[\kappa(1-\theta)+1/4-n]/\delta}}
\over{e\sqrt{\delta^2+\kappa^2}\epsilon}}\
\ \ ; \ \ \ n\in\mathbb{Z}
\end{equation}
for the dimensionless discrete radii of the rapidly-spinning exotic
compact objects which can support the stationary bound-state
configurations of the spatially regular massive scalar fields.

\section{Numerical confirmation}

It is of physical interest to confirm the validity of the compact
analytical formulas (\ref{Eq37}) and (\ref{Eq38}) for the discrete
radii of the exotic compact objects which can support the stationary
bound-state massive scalar field configurations. In Table
\ref{Table1} we present the dimensionless radii
$x^{\text{analytical}}_{\text{c}}(n)$ of the horizonless exotic
compact objects with reflecting Dirichlet boundary conditions as
calculated from the analytically derived formula (\ref{Eq37}). We
also present the corresponding dimensionless radii
$x^{\text{numerical}}_{\text{c}}(n)$ of the rapidly-spinning exotic
compact objects as obtained from a direct numerical solution of the
compact resonance condition (\ref{Eq23}). The numerically computed
roots of the Whittaker function $W_{\kappa,i\delta}(2\epsilon x)$
and its spatial derivative [see the analytically derived resonance
equations (\ref{Eq23}) and (\ref{Eq24})] were obtained directly from
the WolframAlpha: Computational Knowledge Engine.

In the physically interesting $x_{\text{c}}\ll1$ regime
\cite{Notephy} [see Eq. (\ref{Eq3})], one finds a remarkably good
agreement between the approximated discrete radii of the horizonless
exotic compact objects [as calculated from the analytical formula
(\ref{Eq37})] and the corresponding exact radii of the
rapidly-spinning exotic compact objects [as obtained numerically
from the characteristic resonance equation (\ref{Eq23})]. It is
worth emphasizing the fact that the characteristic dimensionless
radii $x^{\text{Dirichlet}}_{\text{c}}(n)$ of the supporting exotic
compact objects, as presented in Table \ref{Table1}, conform to the
analytically derived upper bound (\ref{Eq33}).

\begin{table}[htbp]
\centering
\begin{tabular}{|c|c|c|c|c|c|c|c|c|}
\hline \text{Formula} & \ $x^{\text{Dir}}_{\text{c}}(n=0)$\ \ & \
$x^{\text{Dir}}_{\text{c}}(n=1)$\ \ & \
$x^{\text{Dir}}_{\text{c}}(n=2)$ \ \ & \
$x^{\text{Dir}}_{\text{c}}(n=3)$\ \ & \
$x^{\text{Dir}}_{\text{c}}(n=4)$\ \ & \
$x^{\text{Dir}}_{\text{c}}(n=5)$\\
\hline \ {\text{Analytical}}\ [Eq. (\ref{Eq37})]\ \ &\
0.07998\ \ &\ 0.02449\ \ &\ 0.00750\ \ &\ 0.00229\ \ &\ 0.00070\ \ &\ 0.00022\\
\ {\text{Numerical}}\ [Eq. (\ref{Eq23})]\ \
&\ 0.08690\ \ &\ 0.02503\ \ &\ 0.00754\ \ &\ 0.00230\ \ &\ 0.00070\ \ &\ 0.00022\\
\hline
\end{tabular}
\caption{Composed stationary
exotic-compact-object-massive-scalar-field configurations. We
present the dimensionless discrete radii
$x^{\text{Dirichlet}}_{\text{c}}(n)$ of the rapidly-spinning exotic
compact objects with reflective Dirichlet boundary conditions as
calculated from the analytically derived compact formula
(\ref{Eq37}). We also present the corresponding dimensionless radii
of the horizonless exotic compact objects  as obtained from a direct
numerical solution of the characteristic resonance condition
(\ref{Eq23}). The data presented is for massive scalar field
configurations with angular harmonic indices $l=m=10$, $M\omega=5$,
and $\mu=1.5\omega$ [these field parameters correspond to
$\epsilon=5.59$, $\kappa=-1.12$, and $\delta=2.65$, see Eq.
(\ref{Eq17})]. One finds a remarkably good agreement between the
approximated discrete radii of the rapidly-spinning exotic compact
objects [as calculated from the analytically derived compact formula
(\ref{Eq37})] and the corresponding exact radii of the horizonless
exotic compact objects [as obtained from a direct numerical solution
of the Dirichlet resonance equation (\ref{Eq23})]. Note that the
dimensionless radii $x^{\text{Dirichlet}}_{\text{c}}(n)$ of the
supporting rapidly-spinning exotic compact objects conform to the
analytically derived upper bound (\ref{Eq33}).} \label{Table1}
\end{table}

In Table \ref{Table2} we display the dimensionless discrete radii
$x^{\text{analytical}}_{\text{c}}(n)$ of the rapidly-spinning exotic
compact objects with reflecting Neumann boundary conditions as
calculated from the analytically derived formula (\ref{Eq38}). We
also present the corresponding dimensionless radii
$x^{\text{numerical}}_{\text{c}}(n)$ of the horizonless exotic
compact objects as obtained numerically from the resonance condition
(\ref{Eq24}). Again, in the physically interesting
$x_{\text{c}}\ll1$ regime \cite{Notephy} [see Eq. (\ref{Eq3})], one
finds a remarkably good agreement between the approximated discrete
radii of the rapidly-spinning exotic compact objects [as calculated
from the analytically derived formula (\ref{Eq38})] and the
corresponding exact radii of the horizonless exotic compact objects
[as obtained numerically from the characteristic resonance equation
(\ref{Eq24})]. It is worth pointing out that the characteristic
dimensionless radii $x^{\text{Neumann}}_{\text{c}}(n)$ of the
supporting exotic compact objects, as presented in Table
\ref{Table2}, conform to the analytically derived upper bound
(\ref{Eq33}).

\begin{table}[htbp]
\centering
\begin{tabular}{|c|c|c|c|c|c|c|c|c|}
\hline \text{Formula} & \ $x^{\text{Neu}}_{\text{c}}(n=0)$\ \ & \
$x^{\text{Neu}}_{\text{c}}(n=1)$\ \ & \
$x^{\text{Neu}}_{\text{c}}(n=2)$ \ \ & \
$x^{\text{Neu}}_{\text{c}}(n=3)$\ \ & \
$x^{\text{Neu}}_{\text{c}}(n=4)$\ \ & \
$x^{\text{Neu}}_{\text{c}}(n=5)$\\
\hline \ {\text{Analytical}}\ [Eq. (\ref{Eq38})]\ \ &\
0.13476\ \ &\ 0.04125\ \ &\ 0.01263\ \ &\ 0.00387\ \ &\ 0.00118\ \ &\ 0.00036\\
\ {\text{Numerical}}\ [Eq. (\ref{Eq24})]\ \
&\ 0.16106\ \ &\ 0.04291\ \ &\ 0.01277\ \ &\ 0.00388\ \ &\ 0.00118\ \ &\ 0.00036\\
\hline
\end{tabular}
\caption{Composed stationary
exotic-compact-object-massive-scalar-field configurations. We
present the dimensionless discrete radii
$x^{\text{Neumann}}_{\text{c}}(n)$ of the horizonless exotic compact
objects with reflective Neumann boundary conditions as calculated
from the analytically derived formula (\ref{Eq38}). We also present
the corresponding dimensionless radii of the rapidly-spinning exotic
compact objects as obtained from a direct numerical solution of the
characteristic resonance condition (\ref{Eq24}). The data presented
is for massive scalar field configurations with angular harmonic
indices $l=m=10$, $M\omega=5$, and $\mu=1.5\omega$ [these field
parameters correspond to $\epsilon=5.59$, $\kappa=-1.12$, and
$\delta=2.65$, see Eq. (\ref{Eq17})]. One finds a remarkably good
agreement between the approximated discrete radii of the horizonless
exotic compact objects [as calculated from the analytically derived
compact formula (\ref{Eq38})] and the corresponding exact radii of
the rapidly-spinning exotic compact objects [as obtained from a
direct numerical solution of the Neumann resonance equation
(\ref{Eq24})]. Note that the dimensionless radii
$x^{\text{Neumann}}_{\text{c}}(n)$ of the supporting
rapidly-spinning exotic compact objects conform to the analytically
derived upper bound (\ref{Eq33}).} \label{Table2}
\end{table}

\section{Summary and Discussion}

Some candidate quantum-gravity models
\cite{eco1,eco2,eco3,eco4,eco5,eco6,eco7,eco8,eco9,eco10,eco11,eco12,Pan}
have recently put forward the intriguing idea that quantum effects
may prevent the formation of stable black-hole horizons. These
models have suggested, in particular, that within the framework of a
quantum theory of gravity, horizonless exotic compact objects may
serve as alternatives to classical black-hole spacetimes.

Motivated by this intriguing prediction, we have raised here the
following physically interesting question: Can horizonless exotic
compact objects with reflective boundary conditions \cite{Pan} (see
also
\cite{eco1,eco2,eco3,eco4,eco5,eco6,eco7,eco8,eco9,eco10,eco11,eco12})
support spatially regular massive scalar field configurations in
their exterior regions? In order to address this intriguing
question, in the present paper we have solved {\it analytically} the
Klein-Gordon wave equation for a stationary linearized scalar field
of mass $\mu$, proper frequency $\omega_{\text{c}}$, and spheroidal
harmonic indices $(l,m)$ in the background of a rapidly-spinning
[see Eq. (\ref{Eq14})] exotic compact object of mass $M$ and angular
momentum $J=Ma$. The main physical results derived in the present
paper are as follows:

(1) It was proved that the compact upper bound [see Eqs.
(\ref{Eq17}) and (\ref{Eq33})]
\begin{equation}\label{Eq46}
x_{\text{c}}<{{\kappa+\sqrt{\kappa^2+\delta^2}}\over{\epsilon}}\
\end{equation}
on the dimensionless radius of the spinning exotic compact object
provides a necessary condition for the existence of the composed
stationary exotic-compact-object-massive-scalar-field
configurations.

(2) We have shown that, for a given set $(\mu,l,m)$ of the physical
parameters which characterize the massive scalar field, there exists
a {\it discrete} spectrum of radii $\{r_{\text{c}}(\mu,l,m,M,a;n)\}$
of the rapidly-spinning exotic compact objects which can support the
stationary bound-state massive scalar field configurations. In
particular, it was proved analytically that the compact resonance
conditions [see Eqs. (\ref{Eq23}) and (\ref{Eq24})]
\begin{equation}\label{Eq47}
W_{\kappa,i\delta}(2\epsilon x^{\text{Dirichlet}}_{\text{c}})=0\ \ \
\ ; \ \ \ \ {{d}\over{dx}}[x^{-1}W_{\kappa,i\delta}(2\epsilon
x)]_{x=x^{\text{Neumann}}_{\text{c}}}=0\
\end{equation}
determine the characteristic discrete sets of supporting radii which
characterize the rapidly-spinning reflective exotic compact objects.

(3) It was explicitly shown that the physical properties of the
composed stationary exotic-compact-object-massive-scalar-field
configurations can be studied {\it analytically} in the physically
interesting $x_{\text{c}}\ll1$ regime \cite{Notephy} [see Eq.
(\ref{Eq3})]. In particular, we have used analytical techniques in
order to derive the compact formula [see Eqs. (\ref{Eq37}) and
(\ref{Eq38})]
\begin{equation}\label{Eq48}
x_{\text{c}}(n)={{e^{-\pi
n/\delta}}\over{2\epsilon}}\Big[\nabla{{\Gamma(1+2i\delta)\Gamma({1\over
2}-i\delta-\kappa)}\over{\Gamma(1-2i\delta)\Gamma({1\over
2}+i\delta-\kappa)}}\Big]^{1/2i\delta}\ \ \ \ ; \ \ \ \
\nabla=\begin{cases}
1 &\ \ \ \ \text{Dirichlet B. C.}\  \\
{{1+2i\delta}\over{1-2i\delta}} &\ \ \ \ \text{Neumann B. C.}\ \
\end{cases}
%\ \ \ ; \ \ \ n\in\mathbb{Z}
\end{equation}
for the discrete spectra of reflecting radii
$\{x_{\text{c}}(\mu,l,m,M,a;n)\}$ which characterize the
rapidly-spinning exotic compact objects that can support the
stationary bound-state massive scalar field configurations.

(4) We have explicitly shown that, in the physically interesting
$x_{\text{c}}\ll1$ regime \cite{Notephy}, the analytically derived
formulas (\ref{Eq37}) and (\ref{Eq38}) for the discrete radii of the
horizonless exotic compact objects that can support the stationary
bound-state massive scalar field configurations agree with direct
numerical computations of the corresponding radii of the
rapidly-spinning exotic compact objects.

(5) It is worth emphasizing again that the composed
spinning-exotic-compact-object-massive-scalar-field configurations
that we have studied in the present paper, like the composed
spinning-black-hole-massive-scalar-field configurations studied
recently in \cite{Hodrc,Herkr}, owe their existence to the
intriguing physical phenomenon of superradiant scattering in
rotating spacetimes \cite{Frid,Bri}. In particular, the spatially
regular stationary massive scalar field configurations (\ref{Eq22})
are characterized by the critical (marginal) frequency
$\omega_{\text{c}}$ for the superradiant scattering phenomenon of
bosonic (integer-spin) fields in spinning spacetimes [see Eqs.
(\ref{Eq10}) and (\ref{Eq14})].

(6) Finally, we would like to stress the fact that, combining the
results of the present paper with the results presented in
\cite{Hodrc,Herkr}, one deduces that both spinning black holes and
horizonless spinning exotic compact objects with reflecting surfaces
can support spatially regular configurations of stationary massive
scalar fields in their exterior regions. It is important to note,
however, that for given physical parameters $\{M,a\}$ of the central
supporting object, the discrete resonant spectra $\{\mu(M,a)\}$ of
the allowed field masses are different. Thus, our analytically
derived theoretical results may one day be of practical
observational importance since the different resonant spectra of the
external stationary massive scalar field configurations may help
astronomers to distinguish horizonless spinning exotic compact
objects from genuine black holes.

In particular, it is worth pointing out that the regime of existence
of the composed spinning-black-hole-massive-scalar-field
configurations is characterized, in the rapidly-rotating $a/M\to 1$
limit, by the dimensionless relations $m/2<M\mu<m/\sqrt{2}$
\cite{Hodrc}. On the other hand, the composed
spinning-exotic-compact-object-massive-scalar-field configurations
exist for all real values of the physical parameters $\delta$ and
$\kappa$ [see Eq. (\ref{Eq19})]. Taking cognizance of the large-$m$
relation (\ref{Eq41}), one deduces that the horizonless spinning
configurations studied in the present paper have the {\it larger}
regime of existence $m/2<M\mu<\sqrt{3}m/2$.

\bigskip
\noindent
{\bf ACKNOWLEDGMENTS}
\bigskip

This research is supported by the Carmel Science Foundation. I thank
Yael Oren, Arbel M. Ongo, Ayelet B. Lata, and Alona B. Tea for
stimulating discussions.

%\newpage


\begin{thebibliography}{99}
\bibitem{Notennn} Here the integer $n$ is the resonance parameter of
the composed exotic-compact-object-linearized-massive-scalar-field
configurations [see Eqs. (\ref{Eq37}) and (\ref{Eq38}) below].
\bibitem{Noteman} Following \cite{Pan}, we shall assume that the energy and angular
momentum of the reflective quantum membrane are negligible.
\bibitem{Noteunm} Note that the proper mass of the scalar field stands for $\mu/\hbar$.
This characteristic physical parameter therefore has the dimensions
of $($length$)^{-1}$.
\bibitem{Notedec} Here $\{\omega,l,m\}$ are respectively
the characteristic proper frequency, the spheroidal harmonic index,
and the azimuthal harmonic index (with $l\geq |m|$) of the
stationary spatially regular scalar field configuration.
\bibitem{Notehc} Here $M_{\kappa,\beta}(z)$ and $W_{\kappa,\beta}(z)$
are the Whittaker function of the first kind and the Whittaker
function of the second kind \cite{Abram}, respectively.
\bibitem{Notedp} One may assume, without loss of generality, that
$\delta>0$.
\bibitem{Noteasa} Here we have used Eqs. 13.1.4, 13.1.8, 13.1.32,
and 13.1.33 of \cite{Abram}.
\bibitem{Notenn} Here we have used the relation $1=e^{-2i\pi n}$, where the integer
$n$ is the resonance parameter of the composed
exotic-compact-object-massive-scalar-field configurations.
\bibitem{Notekp0} It is worth noting that the analytically derived resonance spectra (\ref{Eq37}) and
(\ref{Eq38}), which characterize the composed
exotic-compact-object-linearized-massive-scalar-field
configurations, can be further simplified in the asymptotic regime
$\omega^2/\mu^2\to 1^-$ of marginally-bound massive scalar field
configurations [see Eq. (\ref{Eq8})]. These marginally-bound
resonances correspond to $\kappa\to\infty$ [see Eq. (\ref{Eq17})],
in which case one may use the asymptotic approximation
$\Gamma(1/2-i\delta-\kappa)/\Gamma(1/2+i\delta-\kappa)=\kappa^{-2i\delta}e^{2i\pi\kappa}[1+O(\kappa^{-1})]$
\cite{Abram}. Substituting this dimensionless ratio into
(\ref{Eq37}) and (\ref{Eq38}), one finds the simplified resonant
spectra
$x^{\text{Dirichlet}}_{\text{c}}(n)={{2e^{\pi(\kappa-n)/\delta}}\over{m^2}}
\Big[{{\Gamma(1+2i\delta)}\over{\Gamma(1-2i\delta)}}\Big]^{1/2i\delta}$
and
$x^{\text{Neumann}}_{\text{c}}(n)={{2e^{\pi(\kappa-n)/\delta}}\over{m^2}}
\Big[{{\Gamma(2+2i\delta)}\over{\Gamma(2-2i\delta)}}\Big]^{1/2i\delta}$
[Here we have used the dimensionless relation $M\omega={1\over
2}m\cdot[1+O(\tau,\tau\varpi)]$ in the regime (\ref{Eq14}) of
rapidly-spinning exotic compact objects]. Note also that
$K_{mm}=m(m+1)+O[a^2(\mu^2-\omega^2)]$ in the marginally-bound
$\omega^2/\mu^2\to 1^-$ regime
\cite{Heun,Fiz1,Teuk,Abram,Stro,Hodasy,Hodpp}, which implies
$\delta_{mm}=(m^2/2-m-1/4)^{1/2}$ [see Eq. (\ref{Eq17})].
\bibitem{Notelmlm} Note that the large-$m$ regime (\ref{Eq39}) corresponds to the
large-frequency and large-mass regimes $M\mu>M\omega\gg1$ [see Eqs.
(\ref{Eq8}), (\ref{Eq10}), and (\ref{Eq14})]. Likewise, the
large-$m$ regime (\ref{Eq39}) corresponds to $\delta\gg1$ and
$\kappa\gg1$ [see Eq. (\ref{Eq17})].
\bibitem{Notewlm} It is worth emphasizing that the asymptotic large-$m$ expression (\ref{Eq40})
for the eigenvalues of the characteristic spheroidal angular
equation (\ref{Eq7}) is valid in the regime
$-a^2(\mu^2-\omega^2)<m^2$ \cite{Hodpp}. Taking cognizance of the
inequality $\omega^2<\mu^2$ [see (\ref{Eq8})], which characterizes
the bound-state resonances of the massive scalar fields, one
immediately concludes that the requirement
$-a^2(\mu^2-\omega^2)<m^2$ is trivially satisfied by the stationary
spatially regular scalar configurations.
\bibitem{Notegmf} Here we have used Eq. 6.1.15 of \cite{Abram},
which implies
$\Gamma(1+2i\delta)/\Gamma(1-2i\delta)=-\Gamma(2i\delta)/\Gamma(-2i\delta)$
and
$\Gamma(2+2i\delta)/\Gamma(2-2i\delta)=\Gamma(2i\delta)/\Gamma(-2i\delta)\cdot[1+O(m^{-1})]$
in the asymptotic large-$m$ regime (\ref{Eq39}).
\bibitem{Notephy} It is worth emphasizing again that the strong inequality $x_{\text{c}}\ll1$
characterizes exotic compact objects whose quantum reflective
surfaces are located in the radial vicinity of the would-be
classical black-hole horizons [see Eq. (\ref{Eq3})]. This
small-$x_{\text{c}}$ regime corresponds to the physically
interesting model of spinning exotic compact objects recently
studied numerically in \cite{Pan} (see also
\cite{eco1,eco2,eco3,eco4,eco5,eco6,eco7,eco8,eco9,eco10,eco11,eco12}).
\end{thebibliography}
\end{document}